# Lesion-Independent Associations Between Thalamic Nuclei Volumes and Information Processing Speed in Multiple Sclerosis

Arshya Pooladi-Darvish[1,2], Heather Rosehart[3], Marina R. Everest[3], Ali R. Khan[1,2], Sarah A. Morrow[3,4]

[1]Department of Medical Biophysics, Schulich School of Medicine & Dentistry, Western University, London, Canada; [2]Centre for Functional and Metabolic Mapping, Robarts Research Institute, London, Canada; [3]Department of Clinical Neurological Sciences, London Health Sciences Centre, London, Canada; [4]Department of Clinical Neurosciences, Foothills Medical Centre, University of Calgary, Calgary, Canada

Corresponding author. E-mail: sarah.morrow@albertahealthservices.ca

**Background:** Cognitive impairment in multiple sclerosis (MS) is driven by both focal inflammation and compartmentalized neurodegeneration, yet the relative effect of lesion-independent thalamic atrophy on information processing speed (IPS) remains unclear.

**Methods:** This retrospective cohort study included 100 participants with MS. Automatic segmentation techniques quantified lesion load and delineated 26 thalamic regions of interest (ROIs). Linear models compared associations between ROI volumes and Symbol Digit Modalities Test (SDMT) performance in lesion-adjusted and unadjusted models.

**Results:** Twenty-one of 26 ROIs showed significant SDMT associations before lesion adjustment; twelve remained significant after adjustment. Lesion-independent associations were observed in the global thalamus, sensory relay nuclei (ventral posterolateral, medial and lateral geniculate), and associative hubs (pulvinar and mediodorsal-parafascicular complex). These processing-associated ROIs exhibited significantly lower lesion-mediated effects (13.4%) than those losing significance after adjustment (34.2%, $p < 0.001$).

**Conclusion:** Our findings suggest that IPS impairment reflects heterogeneous contributions from focal lesion-driven and chronic neurodegenerative pathology, with nucleus-specific phenotyping potentially informing identification of higher risk individuals.

## Introduction

The prevailing view is shifting toward recognizing cognitive impairment in multiple sclerosis (MS) as the product of two parallel yet distinct pathological processes: relapsing disease activity driven by inflammatory lesions, and compartmentalized, chronic neurodegeneration that persists independent of acute inflammatory activity.[1] This distinction has important implications for persistent cognitive decline despite suppression of relapse activity. Numerous MRI and neuropathological studies demonstrate that deep grey matter (DGM) atrophy is strongly linked to lesion-mediated effects, suggesting that much of this atrophy occurs downstream of white matter pathology.[2–7] DGM structures, particularly the thalamus, atrophy earlier and faster than cortical regions and demonstrate strong associations with cognitive outcomes.[8–12]

Global and nucleus-specific thalamic atrophy have been demonstrated to be strong correlates of cognitive slowing in MS.[2,4,13–17] Information processing speed (IPS) is the most frequently impaired cognitive domain in MS[18,19] and demonstrates robust correlations with thalamic structural and microstructural integrity.[1,3,4,6,7] These associations reflect multiple converging mechanisms, including white matter injuries and disconnection, chronic neuroinflammation, and the thalamus's anatomical vulnerability to periventricular and vascular pathology.[3,6,7] Many studies continue to treat the thalamus as a uniform structure, neglecting the anatomical and functional heterogeneity that defines its nuclei.[9–11,20] This gap in the literature necessitates a more rigorous, nucleus-specific analysis to disentangle lesion-mediated versus lesion-independent thalamic degeneration by contrasting their associations with IPS which may, in turn, inform earlier detection of cognitive risk.

In this work, we used automated segmentation to quantify thalamic regions of interest (ROIs) and lesion volumes, examining lesion-adjusted and unadjusted associations with IPS.[21,22]

Differentiating focal lesion-driven effects from other neurodegeneration may elucidate the biological basis of cognitive deficits in MS and its relationship to intrinsic grey matter pathology.[1]

## Methods

This study was approved by the University of Western Ontario (Western University) Health Sciences Review Ethics Board (HSREB #11520). All participants provided written informed consent prior to any study procedures taking place. Financial support was provided by an investigator-initiated grant from Biogen Idec Canada.

### Participants

Participants with clinically definite MS aged 18 and older were recruited from the London (ON) MS clinic (London, ON Canada) between November 2020 and August 2022. Participants were excluded if they had binocular vision worse than 20/70, had relapsed within 90 days, had changed any medication within 30 days, or had an Expanded Disability Status Scale (EDSS) greater than or equal to 8.0.

### Clinical Measures

The oral Symbol Digit Modalities Test (SDMT) was administered by a trained research coordinator.[23] The SDMT is a validated measure of IPS where participants are presented with a 8½ × 11 inch sheet of paper with nine symbols paired with corresponding digits (1–9) at the top of the page. Below this, the page contains a randomized, sequential assortment of these symbols. Participants are asked to verbally indicate the correct number for each symbol while the coordinator denotes their replies. Participants' score is the total number of correct responses in 90 seconds.

## MRI

Participants were scanned on a 3 Tesla Siemens MAGNETOM Prisma Fit whole-body scanner using a 32-channel head coil. T1-weighted (T1w) images were acquired using a 3D magnetization prepared rapid gradient echo (MPRAGE) sequence (repetition time (TR) = 2400 ms, echo time (TE) = 2.28 ms, flip angle = 9º, voxel size = 0.8 mm isotropic, GRAPPA acceleration factor R = 2). A 3D T2-weighted fluid-attenuated inversion recovery (FLAIR) was acquired with TR = 5000 ms, TE = 387 ms, TI = 1800 ms, flip angle = 120°, voxel size = 0.8 mm isotropic.

## Image Preprocessing

Lesion segmentation was performed using the Lesion Segmentation Toolbox's LST-AI.[22] LST-AI bias-corrected and co-registered T1w and FLAIR images and applied an ensemble of three 3D U-Net convolutional neural networks trained on a large MS dataset (n = 491) and further validated on 103 independent cases.[22,24] The resulting lesion probability masks were then thresholded to produce individualized binary lesion masks in participants' native space. Lesion load was defined as the total lesion volume and log-transformed to correct for right skew.

Thalamic nuclei were segmented using HIPS-THOMAS which extends the Thalamus Optimized Multi-Atlas Segmentation (THOMAS) framework to support MPRAGE images.[21] Following bias-correction, synthetic white-matter-nulled (WMn) images were generated to enhance grey-white matter contrast. WMn images were non-linearly registered to the THOMAS template, and the resulting atlas labels were inverse warped into each participant's native space to obtain segmentations. This pipeline yielded 13 thalamic ROIs per hemisphere, including the global thalamus, 9 major nuclei (anteroventral, ventral anterior, ventral lateral anterior, ventral lateral posterior, ventral posterolateral, central medial, pulvinar, lateral geniculate, and medial geniculate), two perithalamic structures (habenula and mammillothalamic tract), and a single ROI representing the mediodorsal-parafascicular (MD-Pf) complex (Supplementary Figure 1).

## Statistical Analysis

Descriptive statistics were used to summarize the demographic, clinical, and imaging variables. There was no missing cognitive or imaging data. The raw SDMT score served as the dependent variable. For each ROI, the THOMAS-derived volume was normalized to intracranial volume (ICV) to control head-size variability and was *z*-scored across all participants. All analyses were performed separately for left and right hemispheres to explore lateralization of effects.

Analysis of covariance (ANCOVA) models were fit with ROI volume as the independent variable, and age, sex, and years of education were included as covariates. A second set of models evaluated lesion-independent associations by adding lesion load as an additional covariate. Because ROI volumes were *z*-scored, the resulting $\beta$ coefficients represent the expected change in SDMT per one standard deviation change in ROI volume. Multiple comparisons across nuclei were controlled using the Benjamini-Hochberg procedure, and false discovery rate (FDR)-adjusted $p < 0.05$ was considered significant.

Post hoc analysis further examined whether the associations between ROI volume and SDMT performance were mediated or explained by lesion burden. Single-level mediation models estimated the indirect effect of lesion load using bootstrap resampling with 5000 iterations. Additionally, nested ANCOVA models were compared to determine the change in explained variance ($\Delta R^2$) attributable to adding lesion load as a covariate. Group differences in mediation proportions and $\Delta R^2$ between ROIs that remained significant after lesion adjustment and those significant only in the unadjusted model were assessed using Student's *t*-test.

## Results

Participants were on average 46.2 ± 12.4 years old with a mean disease duration of 11.0 ± 8.8 years (Table 1). SDMT performance averaged 59.2 ± 10.9 points while median EDSS was 2.0 (range: 0.0–7.0).

**Table 1:** Demographic, clinical, and imaging characteristics of participants.

| | **N = 100** |
|---|---|
| Age in years, mean (SD) | 46.2 (12.4) |
| Sex, female, N (%) | 75 (75%) |
| Education in years, mean (SD) | 14.2 (2.3) |
| Disease duration, mean in years (SD) | 11.0 (8.8) |
| SDMT, mean (SD) | 59.2 (10.9) |
| EDSS, median (range) | 2.0 (0.0–7.0) |
| Total lesion volume in mL, mean (SD) | 7.0 (6.8) |
| Intracranial volume in mL, mean (SD) | 1582 (150) |

### Lesion-Dependent Associations

In the models excluding lesion load as a covariate, 21 of the 26 thalamic ROIs were statistically significant after FDR correction. All significant correlations were positive, indicating that smaller ROI volume was associated with lower SDMT scores (Supplementary Figure 2). In the left hemisphere, the strongest associations ($p < 0.001$) were observed for the global thalamus (THAL), pulvinar (Pul), ventral posterolateral (VPL), medial geniculate nucleus (MGN), lateral geniculate nucleus (LGN), and MD-Pf complex. Significant associations ($p < 0.01$) were detected in the central medial (CM) nucleus, with further associations ($p < 0.05$) in the anteroventral (AV), ventral anterior (VA), ventral lateral posterior (VLP) and ventral lateral anterior (VLa) nuclei. In the right hemisphere, significant associations ($p < 0.001$) were observed for THAL, Pul, and MD-Pf complex, with further significant effects ($p < 0.01$) in the VLa, VA, VLP, VPL, MGN, and LGN, and ($p < 0.05$) in the AV nucleus.

## Lesion-Independent Associations

Lesion adjustment isolated 12 ROIs whose volumes remained significantly associated with SDMT after FDR correction (Figure 1). As in the previous model, all surviving correlations were positive. In the left hemisphere, THAL, VPL, Pul, and MGN demonstrated the largest correlations ($p < 0.01$) with the LGN and MD-Pf complex remaining significant but with smaller effects ($p < 0.05$). In the right hemisphere, significant associations ($p < 0.01$) were observed for THAL, with further significant correlations in VA, VLa, Pul, LGN, and MD-Pf complex ($p < 0.05$). β was generally larger across left-hemisphere ROIs, most notably within the pulvinar nucleus (left: $\beta = 4.79$; right: $\beta = 3.49$).

## Post-hoc analysis

The post-hoc analysis further quantified the contributions of lesion-mediated and independent thalamic atrophy to SDMT. Twelve ROIs remained significant after adjusting for lesion load. To characterize these two groups in accordance with their distinct anatomical and functional profiles, we classified the 12 ROIs retaining lesion-independent SDMT associations as processing-associated nuclei, reflecting their direct involvement in sensory relay and higher-order integrative circuits that support information processing speed. The remaining 9 ROIs, whose SDMT associations did not survive lesion adjustment, were classified as tract-mediated nuclei, consistent with their reliance on white matter projections to frontal motor and cingulate cortices and the predominance of secondary degeneration in driving their cognitive associations.

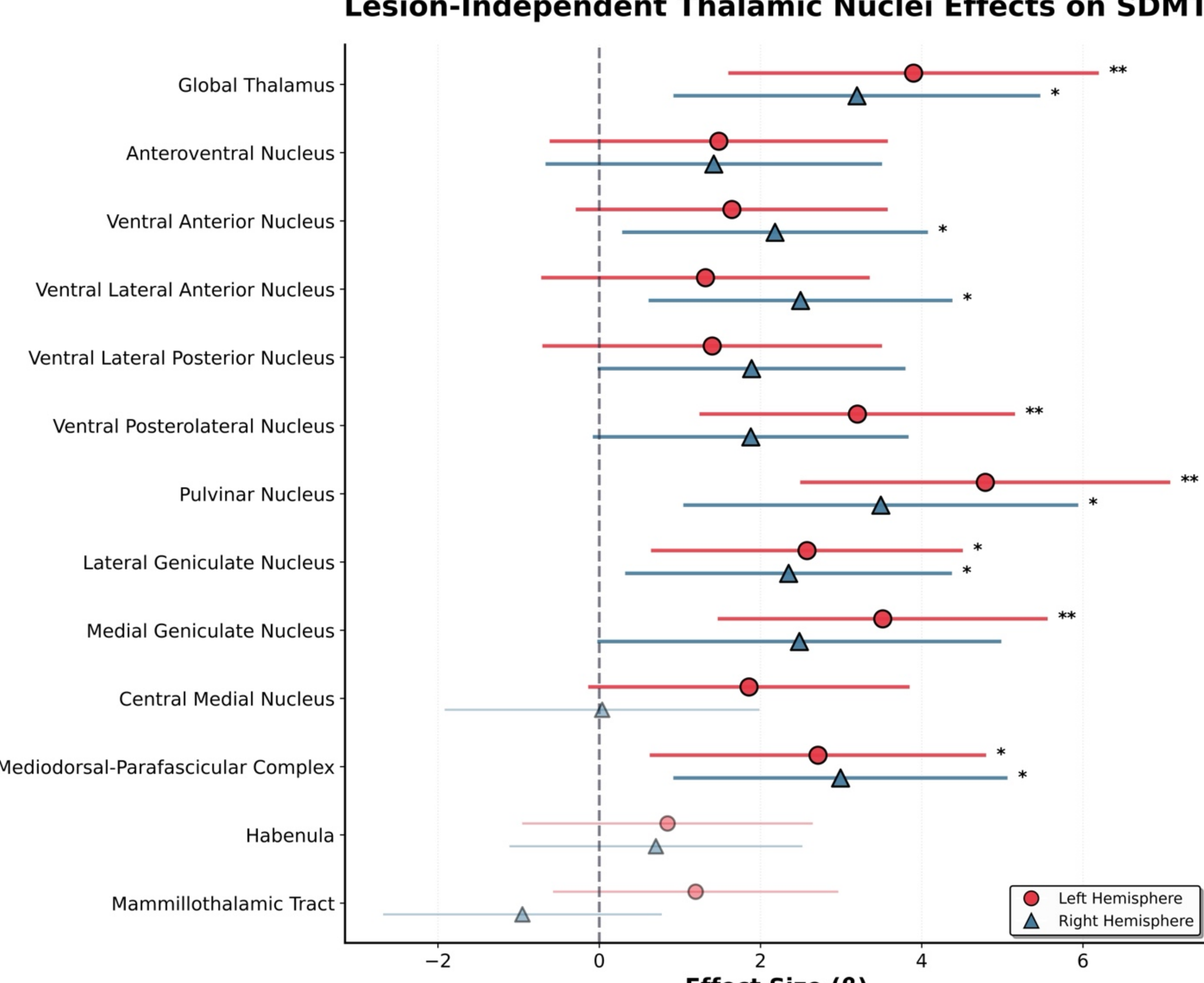

**Figure 1:** Coefficient plot of standardized regression coefficients (β) and 95% confidence intervals for the association of each ROI and SDMT performance after adjusting for age, sex, education, and lesion load. β coefficients reflect the change in SDMT per standard deviation increase in ROI volume. Left hemisphere nuclei are shown as circles; right hemisphere nuclei as triangles. Thicker lines indicate significant associations in the baseline models, asterisks denote ROIs significantly associated after lesion adjustment (*FDR-adjusted $p < 0.05$, **$p < 0.01$).

Mediation analysis revealed that lesion load accounted for a significantly larger proportion of the association between ROI volume and SDMT in tract-mediated nuclei (34.2%) than in processing-associated nuclei (13.4%, $p < 0.001$; Figure 2A), indicating that focal lesion burden plays a substantially greater role in the cognitive associations of tract-mediated structures. Similarly, adding lesion load as a covariate explained a greater share of additional SDMT variance for tract-mediated nuclei ($\Delta R^2 = 5.5\%$) than for processing-associated nuclei ($\Delta R^2 = 2.4\%$, $p < 0.01$; Figure 2B).

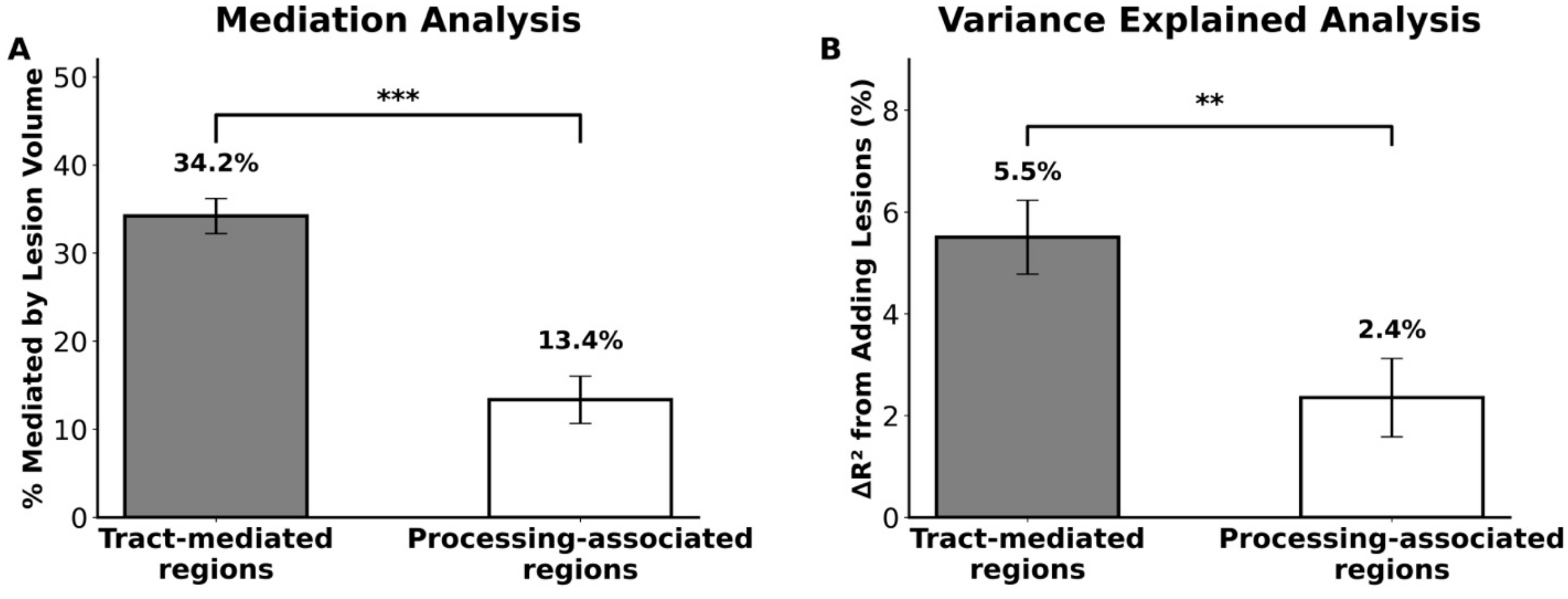

**Figure 2:** Contribution of lesion load to the association between ROI volume and SDMT in tract-mediated and processing-associated ROIs. (A) Proportion of the association between ROI volume and SDMT mediated by lesion load. (B) Additional variance in SDMT performance ($\Delta R^2$) explained by lesion load. Error bars represent standard error. $^{**}p < 0.01$, $^{***}p < 0.001$.

## Discussion

In this work, we contrasted lesion-adjusted and unadjusted models to characterize lesion-mediated and lesion-independent thalamic atrophy associated with deficits in IPS. Nearly half of all thalamic ROIs maintained significant associations with SDMT after adjusting for lesion load, consistent with a heterogeneous contribution of lesion-mediated and lesion-independent factors to thalamic atrophy. While prior studies have correlated nucleus-specific atrophy and cognitive deficits,[6,7,25–27] lesion effects are rarely accounted for. One prior study contrasted models with and without lesion load as a covariate,[4] but that analysis relied on semi-manual lesion segmentation and did not find significant correlations when adjusting for ICV and lesion load. With fully automated segmentation and ICV normalization, our analysis identified significant lesion-independent associations in twelve thalamic ROIs.

We observed nucleus-specific patterns that aligned with distinct thalamocortical circuits. The processing-associated ROIs were generally comprised of first-order sensory relays (VPL, LGN, MGN), and higher-order associative hubs (Pul, MD-Pf). The tract-mediated ROIs were generally linked to motor relay (VA, VLa, VLP), limbic function (AV), and arousal regulation (CM), with projections predominantly to frontal motor and cingulate cortices.[21,28] As the tract-mediated associations with SDMT lost significance after lesion adjustment, lesion-driven degeneration may be a more substantial driver of IPS deficits in these structures. We also noted a modest left-hemisphere dominance across several ROIs. Although this lateralization is subtle, the pattern is in line with the left hemisphere's specialization for symbolic mapping and rule-based attentional control.[29]

Several mechanisms may explain the vulnerability of these processing-associated ROIs to lesion-independent atrophy. First-order sensory nuclei exhibit tonic relay activity, and higher-order associative nuclei engage in sustained cortico-thalamic communication, potentially exerting substantial metabolic demands.[30,31] Within the chronically inflamed MS brain, even slight disruptions to mitochondrial efficiency or axonal energy supply may have disproportionate effects on neuronal stability.[32] Sensory relay and higher-order nuclei, which fire continuously, may face greater oxidative stress than motor relay nuclei, which fire in brief, movement-locked bursts.[33] Their participation in long-range thalamocortical loops may increase dependence on oligodendrocyte support and susceptibility to microglial-mediated injury.[5,7,32] Additionally, thalamic nuclei adjacent to the third ventricle, including the MD-Pf complex, exhibit accelerated neurodegeneration and an *ependymal-in* gradient of microglial activation.[6,26] Trans-synaptic degeneration may also contribute to thalamic atrophy as corticothalamic projections can propagate degenerative signals anterogradely or retrogradely.[3,13] That these ROIs exhibit lesion-independent SDMT associations suggests that their atrophy is not mainly a downstream consequence indicating acute inflammation, but rather reflects larger contributions from chronic neurodegeneration consistent with smouldering disease processes.[1,7]

Crucially, these same nuclei subserve the cognitive operations that the SDMT is sensitive to.[19] The VPL, MGN, and LGN provide principal somatosensory and visual relays to the cortex, while Pul and MD-Pf regulate corticocortical communication and attentional control during goal-directed behaviour.[28] Collectively, these ROIs form the neural basis of rapid sensory encoding, attentional selection, and visuoperceptual integration that underpin IPS,[18,34] and their atrophy may contribute to cognitive slowing because these nuclei occupy functionally critical positions within IPS-related circuits. Notably, both processing associated and tract-mediated nuclei exhibited comparable SDMT associations in unadjusted models, suggesting that clinical utility of such a distinction lies in identifying distinct pathological substrates.

Our findings highlight the potential utility of these processing-associated thalamic ROIs as markers of cognitive dysfunction, offering greater specificity than global thalamic volume alone. With increasingly accessible automated pipelines for thalamic nuclei segmentation, such measures could screen individuals at greater risk for cognitive decline and serve as endpoints in neuroprotection trials.[21] Regional atrophy is regarded as paraclinical evidence of progression,[35] yet current guidelines do not specify which structures best capture cognitive risk. This distinction has direct clinical relevance: high-efficacy disease-modifying therapies (HE-DMTs) effectively suppress relapse activity, but cognitive decline persists in many individuals despite inflammatory control.[1] Future trials evaluating these agents may benefit from assessing whether protection of these thalamic nuclei preserves IPS.

This study has several limitations. The cross-sectional design prevents inferences about the temporal relationship of thalamic atrophy and cognitive decline. The cohort was recruited from one centre and exhibited mild disability which limits generalizability. Cognitive assessment was restricted to the SDMT and did not consider other domains such as memory, executive function, and language. Lesion burden was derived from total whole brain lesion volume rather than thalamic or tract-specific lesion loads, without accounting for spatial localization. This precludes assessment of whether focal lesions differentially affect tract-mediated versus processing-associated nuclei through local tissue damage. The use of whole brain lesion volume as a covariate nonetheless provides a conservative test of lesion-independent associations as ROIs retaining significance after this adjustment demonstrate IPS associations beyond what focal lesion burden can account for. While all segmentations were manually reviewed, characterization of thalamic nuclei may have been hindered by partial volume effects and misclassification. Due to our sample size and statistical approach, this analysis may have limited power to capture nonlinear or subtler nucleus-specific effects. Nonetheless, this work is the first to demonstrate that lesion adjustment selectively abolishes IPS associations for some thalamic nuclei but not others. We anticipate that these findings will motivate longitudinal imaging studies that probe these mechanisms directly, both within the thalamus and across other regions that may exhibit similar trajectories.

In summary, this study differentiated lesion-dependent and lesion-independent thalamic contributions to IPS deficits in MS. Twelve thalamic ROIs showed significant associations with SDMT performance independent of lesion load. These nuclei exhibited significantly lower lesion-mediated effects than the ROIs whose associations were no longer significant after accounting for lesion load, consistent with a framework involving contributions from focal lesion-driven and chronic neurodegenerative pathology. This nucleus-resolved separation of lesion-independent and lesion-dependent effects motivates further investigation of these mechanisms across vulnerable brain regions.

## Declaration of Conflicting Interests

Arshya Pooladi-Darvish, Heather Rosehart, Marina Everest, and Ali R. Khan report no disclosures relevant to the manuscript. Sarah A. Morrow has served on advisory boards for Biogen Idec, Celgene/BMS, EMD Serono, Novartis, Roche, Sanofi Genzyme, and Teva Neurosciences; received investigator-initiated grant funds from Biogen Idec, Novartis, Roche, and Sanofi Genzyme; and acted as site PI for multicenter trials funded by Celgene/BMS, EMD Serono, Novartis, Roche, and Sanofi Genzyme.

## Funding

Financial support was provided by an investigator-initiated grant from Biogen Idec Canada.

## Data Availability Statement

The data that support the findings of this study are available from the corresponding author upon reasonable request.

## Supplementary Materials

Supplementary material for this article is available online. Supplementary Figure 1 shows a representative thalamic nuclei segmentation using HIPS-THOMAS, displayed in orthogonal views with three-dimensional surface rendering across all 12 segmented structures. Supplementary Figure 2 presents a forest plot of standardized regression coefficients for the association between individual thalamic nuclei volumes and SDMT performance, adjusted for age, sex, and education, with left and right hemisphere results shown separately.

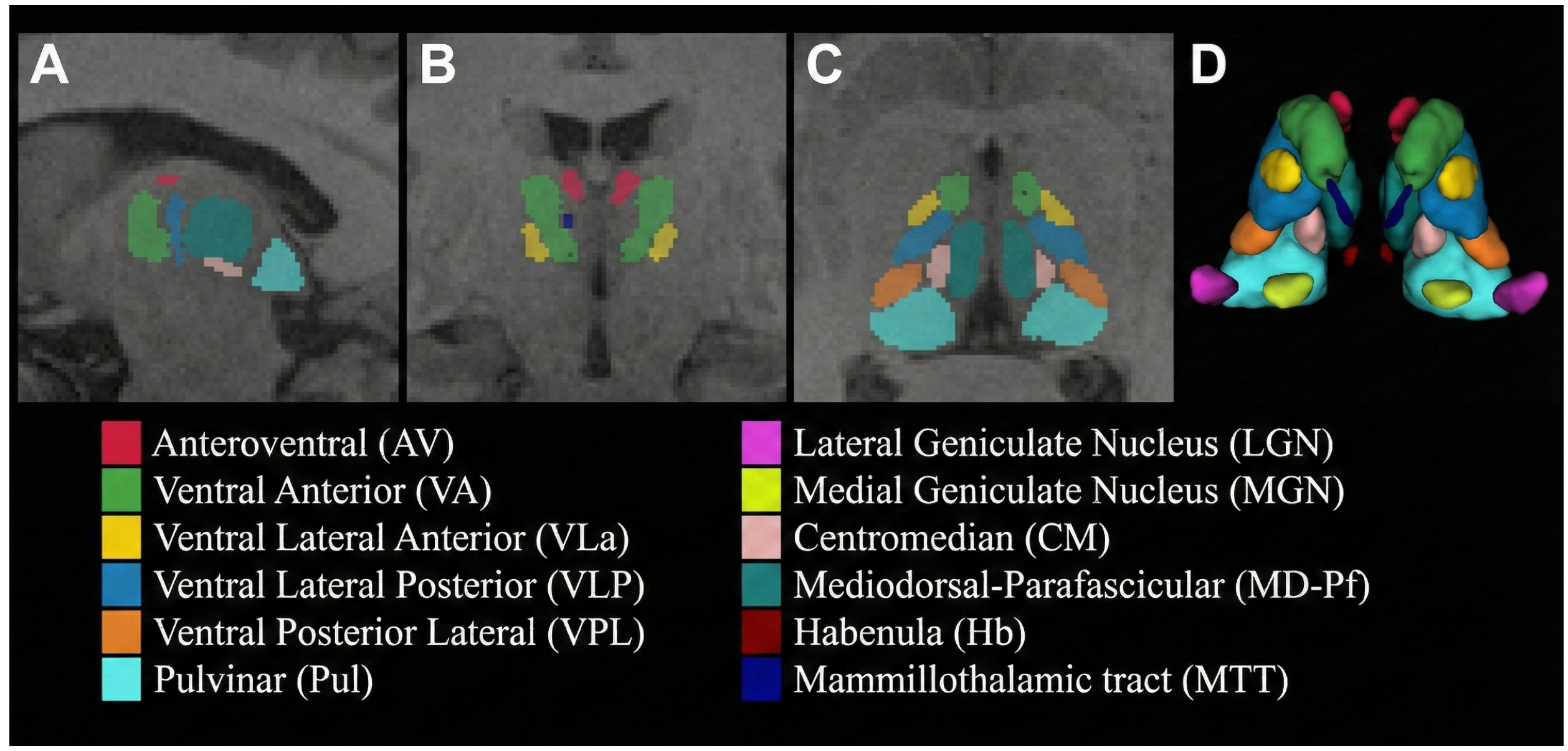

**Supplementary Figure 1. Thalamic nuclei segmentation using HIPS-THOMAS.** Representative segmentation of bilateral thalamic nuclei in a participant with multiple sclerosis. Orthogonal views show the T1-weighted MPRAGE image with segmentation overlay in (A) sagittal, (B) coronal, and (C) axial planes. (D) Three-dimensional surface rendering. Twelve thalamic structures were segmented: anteroventral (AV), ventral anterior (VA), ventral lateral anterior (VLa), ventral lateral posterior (VLP), ventral posterior lateral (VPL), pulvinar (Pul), lateral geniculate nucleus (LGN), medial geniculate nucleus (MGN), centromedian (CM), mediodorsal-parafascicular (MD-Pf), habenula (Hb), and mammillothalamic tract (MTT).

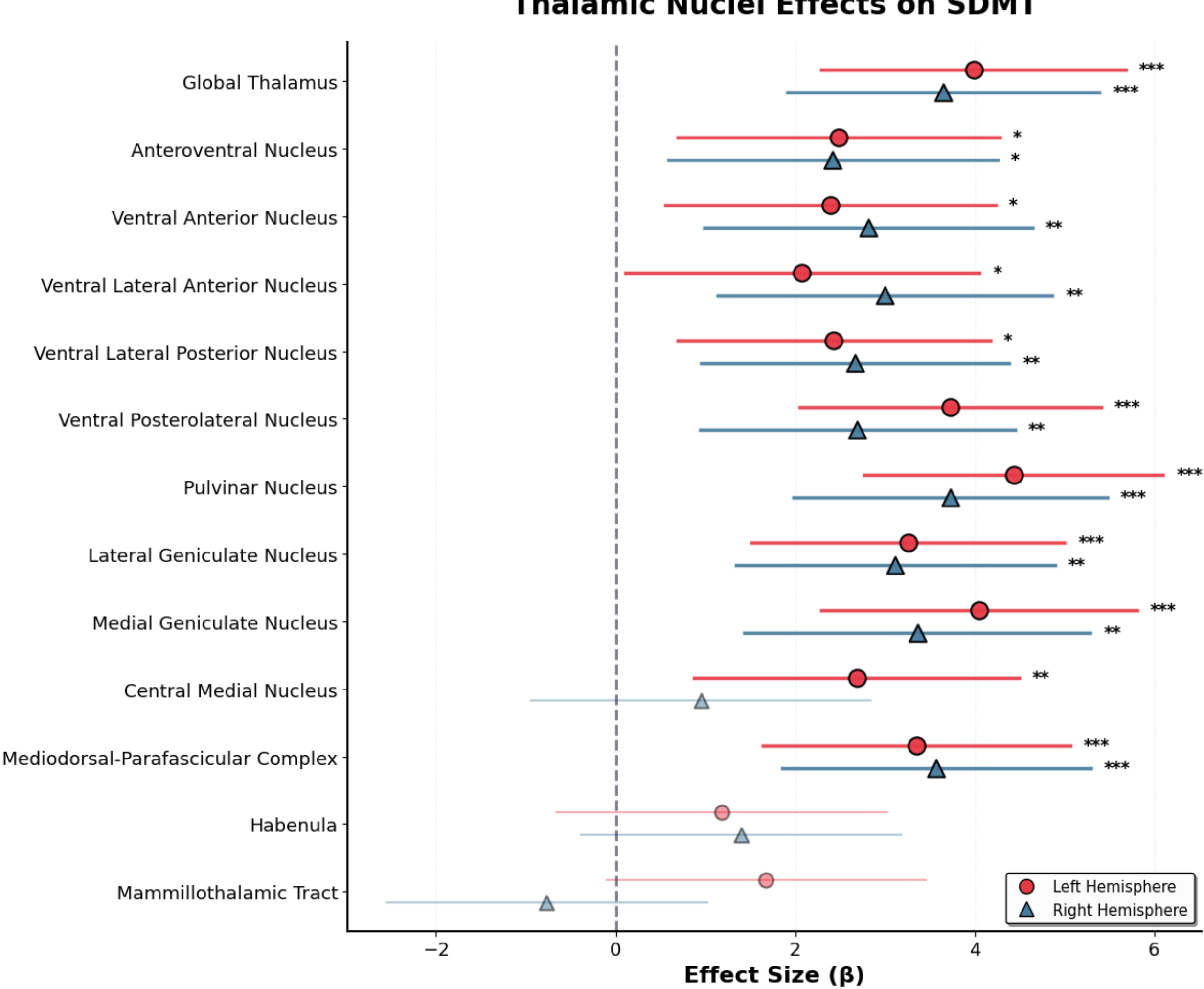

**Supplementary Figure 2. Association between thalamic nuclei volumes and information processing speed.** Forest plot showing standardized regression coefficients (β) for the relationship between individual thalamic nuclei volumes and Symbol Digit Modalities Test (SDMT) performance. Models were adjusted for age, sex, and education. Left hemisphere nuclei are shown as circles; right hemisphere nuclei as triangles. Error bars represent 95% confidence intervals. Asterisks indicate statistical significance: *FDR-adjusted $p < 0.05$, **$p < 0.01$, ***$p < 0.001$.